# The financial framework of the sustainability of health universal coverage in Italy

A quantitative financial model for the assessment of the italian stability and reform program of public health financing

Stefano Olgiati PhD[1], Prof Alessandro Danovi MSc[2]

with

Achille Lanzarini MA[3], Prof Giancarlo Cesana MD[4,5]

[1]*CO-AUTHOR - Ricercatore – Centro Sanità Pubblica – Università degli Studi di Milano-Bicocca - Milano*
[2]*CO-AUTHOR - Professore Associato di Economia e Gestione delle Imprese - Università degli Studi di Bergamo – Bergamo*
[3]*Ricercatore – Centro Sanità Pubblica – Università degli Studi di Milano-Bicocca - Milano*
[4]*Professore Ordinario di Igiene Generale e Applicata - Università degli Studi di Milano-Bicocca- Milano*
[5]*Direttore - Centro Sanità Pubblica – Università degli Studi di Milano-Bicocca- Milano*

# Abstract

**Background**

*Context*

Italy and the Eurozone are heading in the year 2012 into a financial depression of unprecedented magnitude, with a forthcoming multitude of often contradictory public economic and financial stability emergency interventions whose ultimate endogenous and exogenous effects on public and private health spending and on the sustainability of universal coverage are difficult to predict *ex ante*;

*Aim*

The research question is to assess whether it is possible to synthesise into a single and simple quantitative index such multitude of emergency public economic and financial stability interventions and assess their magnitude and direction towards increasing or decreasing sustainability of publicly funded health care and universal coverage. Furthermore, given the strong discontinuities in public accounts data and the necessity for rapid *ex ante* analysis and response, whether such synthetic index can utilize readily available on-time data from EUROSTAT and OECD. Finally, to assess utilizing such index the effects on the sustainability of universal health coverage of the Italian Economic and Stability Reform Program 2011-2014;

**Methods**

*Study Design*

We have analyzed the Italian Economic and Stability Reform Program 2011-2014, which incorporates EUROSTAT and OECD health and economic data and guidelines, and we have proposed a quantitative synthetic *sustainability index* $\sigma$ based on simple partial and absolute differential equations. The strength and statistical significance of trends and data from EUROSTAT and OECD have been tested with simple correlation coefficients;

***Analysis:*** The Italian Economic and Stability Reform Program 2011-2014 rests on the basic assumptions that: i) the Gross Domestic Product will grow in the period 2011-2014 exponentially at a rate of 3.18% vs. a growth of 3.29% in the period 1997-2010; ii) the Public Component of Health Care Financing will not grow in the period 2011-2014 vs. a growth of 0.82% in the period 1997-2010; iii) Non-Health Applications growing expenditure - such as sovereign debt service, pensions, unemployment subsidies, sovereign debt fiscal compact and deficit reduction - are not expected to affect health financing and universal coverage;

**Findings**

The *sustainability index* σ highlights that in case the growth of the GDP in the period 2011-2014 be insufficient - as is already the case in the first semester of 2012 - all the assumptions on which the Italian Economic and Stability Reform Program 2011-2014 rests will fall, and Universal Coverage will become unsustainable. In fact the Sustainability Coefficient σ, which lies in the financial sustainability range (*-1<σ<0* ) under the original assumptions of the Program, would grow above 1 ($\sigma \gg 1$) well into the Financial *and Fiscal Unsustainability* range;

**Relevance and Interpretation**:

Health and Public Health professionals should intervene immediately with Italian and Eurozone national budgets planners and financial health regulators before unselective exogenously induced health financing and provision shortages produce irreparable epidemiological effects.

# Introduction

Ashley M Croft and Joanne V Palmer in *Exercise and life expectancy* (Ashley M Croft 2012) observed quite wittily that through increased daily exercise the risk of mortality can be postponed, but it cannot certainly be reduced or eliminated: the benefits of exercise are relative but unfortunately the risk of mortality is still an absolute[1]. We argue in this paper that the same holds true for the growth of public health spending. Just how much growth, for how long and how much expenditure is sustainable is a relative notion, but with a difference from the benefits of daily exercise: in public health financing if you run faster you arrive sooner. The deepening of the Eurozone financial crisis is requiring all policy makers to revise public health spending forecasts and budgets (OECD 2012). The word *sustainability* is continually evoked, more often than not, as a justification for gross cutbacks on expenditure than as a guarantee to a common effort towards an integrated approach to health planning (Russel L Gruen 2008). We argue in this paper that even at such a macro level as national health financial planning, before plunging into micro programming and detail management, some key issues regarding the *financial framework of sustainability* are not addressed comprehensively by policymakers. We devised a very simple *sustainability index* which synthesises into one indicator the unstable dynamics and tradeoffs of some key macro variables which condition all the health-programming micro processes thereafter in an effort to improve the transparency to the approach to health spending and guarantee that such values as universal health coverage be not lost among the complex dynamics of public accounts stability plans. We identified the variables trying to use easily available data form EUROSTAT, ISTAT and OECD and we utilized the *sustainability index* for a practical assessment of the Italian Economic and Financial Stability Plan 2011-2014, also with the scope of exemplification.

---

[1] For the prophets included (Joanne V Palmer Ashley M Croft, «Exercise and life expectancy,» *Lancet*, 3 March 2012: 800.)

# 1. The Exponential Growth of Italian Public Health Financing in the Period 1997-2010

1.1. The Italian Health Care System (FSN) has been reformed in the period 1992-1999[2], following the full implementation of the law 833/78, after which the State, through its publicly tax funded[3] Fondo Sanitario Nazionale (FSN), became the central player in providing health care and guaranteeing universal coverage (Cesana 2005). Not unexpectedly, the Public component of Health Care Financing, as percent of the Total Health Financing[4], has been growing from 70.8% in 1997 to 77.6% in 2010[5] (Correlation Coefficient 0.96, Exponential Growth Rate 0.82%). The Uninsured Private Out-of-Pocket component which though declining still remains very high and consequently a potential source of social instability in times of pervasive economic recession has been exponentially declining from 26.4% to 19.6% (Correlation Coefficient 0.94, Exponential Decline Rate -2.63%). The Residual Health Financing has remained virtually unchanged (Correlation Coefficient not significant), highlighting the residual nature of Privately Insured Health Financing in Italy. (Exhibit 1)

**Exhibit 1**

| THE EXPONENTIAL GROWTH OF PUBLIC HEALTH FINANCING AS PERCENT OF TOTAL FINANCING IN ITALY IN THE PERIOD 1997-2010 | YEAR | 1997 | 2010 | GROWTH AS A FUNCTION OF TIME | | |
|---|---|---|---|---|---|---|
| | PERIODS | 0 | 13 | CORRELATION COEFFICIENT | | EXPONENTIAL GROWTH RATE - LOG REGRESSION |
| Italian Public Health Financing as Percent ofTtotal Health Financing | P | 70.80% | 77.60% | 0.96 | p | 0.82% |
| Italian Out-of-Pocket Health financing | O | 26.40% | 19.60% | -0.94 | o | -2.63% |
| Italian Insured Health financing | I | 2.80% | 2.80% | -0.18 | i | n.m.* |
| Italian Total Health financing | P+O+I | 100.00% | 100.00% | | | |

*n.m. = not meaningful

*SOURCE: Authors' elaborations on OECD HEALTH DATA 2011*

[2] In our research the base year of the reform, whose successive implementations span from 1992 (D.Lgs.502/92) to 1999 (D.Lgs.229/99), will be considered the year 1997, when in Italy Region Lombardia, with the law L.R. 31/97, was the first to fully implement the guidelines of the law 833/78.
[3] The reformed Italian Health Care System (FSN) is funded by direct and indirect taxation. Public funds are pooled centrally and regionally. Centrally pooled funds are allocated to Regional Health Care Systems (FSR) via risk adjusted capitation.
[4] http://www.oecd.org/document/16/0,3746,en_2649_37407_2085200_1_1_1_37407,00.html (accessed 8/18/09)
[5] As a purely indicative term of comparison, the percent of public health Financing on total health Financing is 76.9% in Germany in the year 2009 OECD, *OECD HEALTH DATA 2011* (PARIS: OECD, 2011). http://stats.oecd.org/Index.aspx?DataSetCode=SHA (accessed 8/18/09)

1.2. Among the goals of the health reform were[6] the containment of rising costs and the improvement of the efficiency and effectiveness of public health care provision for all (Manzoli, Villari and Boccia 2008). However, since the full implementation of the reform in 1997, the total Health Financing Propensity [Appendix A] has been growing from 7.70% of the Italian Gross Domestic Product (GDP) in 1997 to 9.60% in 2010 (Correlation Coefficient 0.97, Exponential Growth Rate 1.69%). (Exhibit 2)

1.3. The combined effect of the growing Health Financing Propensity *and* the growing GDP (Correlation Coefficient 0.98, Exponential Growth Rate 3.23%) has fuelled the growth of Total Health Financing at an exponential rate of 4.92% in the same period (Correlation Coefficient 1.00)[7]. (Exhibit 2)

**Exhibit 2**

| THE EXPONENTIAL GROWTH OF THE PUBLIC HEALTH FINANCING PROPENSITY IN ITALY IN THE PERIOD 1997-2010 AND ITS PROJECTION TO THE YEARS 2025 AND 2050 | | OECD ACTUAL | | | | 1997-2010 | | | PROJECTED 2010 - n(t) | | | |
|---|---|---|---|---|---|---|---|---|---|---|---|---|
| | YEAR | 1997 | | 2010 | | GROWTH AS A FUNCTION OF TIME | | | 2025 | | 2050 | |
| | PERIODS | 0 | | 13 | | CORRELATION COEFFICIENT | | EXPONENTIAL GROWTH RATE - LOG REGRESSION | 28 | | 53 | |
| **Gross Domestic Product** | Y | 1,048,766 | | 1,548,816 | | 0.98 | g | 3.23% | 2,515,590 | | 5,645,555 | |
| **Italian Public Health financing Propensity** | Pφ | 5.45% | | 7.45% | | 0.98 | f+p | 2.51% | 10.86% | | 20.34% | |
| Italian Out-of-Pocket Health financing Propensity | Oφ | 2.03% | | 1.88% | | -0.74 | f+o | -0.94% | | | | |
| Italian Insured Health financing Propensity | Iφ | 0.22% | | 0.27% | | 0.46 | f+i | 1.20% | | | | |
| Italian Total Health financing Propensity | φ | 7.70% | | 9.60% | | 0.97 | f | 1.69% | | | | |
| **Italian Public Health financing** | PH | 57,175 | 70.80% | 115,381 | 77.60% | 0.99 | f+p+g | 5.74% | 273,118 | 84.74% | 1,148,278 | 92.18% |
| Italian Out-of-Pocket Health financing | OH | 21,319 | 26.40% | 29,143 | 19.60% | 0.99 | f+o+g | 2.29% | 41,089 | 12.75% | 72,844 | 5.85% |
| Italian Insured Health financing | IH | 2,261 | 2.80% | 4,163 | 2.80% | 0.83 | f+i+g | 4.43% | 8,093 | 2.51% | 24,506 | 1.97% |
| Italian Total Health financing | H | 80,755 | 100.00% | 148,686 | 100.00% | 1.00 | f+g=h | 4.92% | 322,301 | 100.00% | 1,245,629 | 100.00% |

*SOURCE: Authors' elaborations on OECD HEALTH DATA 2011*

1.4. In synthesis, in the period 1997-2010, following the reform of the Health Care System, Italy has increased *both* its propensity to spend in total health care *and* the percent of such spending publicly financed. The result is that the growth of public health financing has exceeded the growth rate of the GDP by as much as 1.8 times. (Exhibit 2 & 3)

---

[6] The Reform of 1992-1999 focused mainly on Hospital Care provision and payment also with the introduction of a prospective Diagnosis Related Group (DRGs) public reimbursement system (tariff-based) in substitution of the preceding retrospective reimbursement system (cost-based).

[7] In the equations of the present paper *inflation* is *invariant* in all the evaluations, since if THE is the Total Health Financing and GDP is the Gross Domestic Product: THE as % of GDP $= \frac{\text{THE} \times (1+\text{inflation rate})}{\text{GDP} \times (1+\text{inflation rate})} \times 100 = \frac{\text{THE}}{\text{GDP}} \times 100$ then THE $= \frac{\text{THE as \% of GDP}}{100} \times \text{GDP}$

**Exhibit 3**

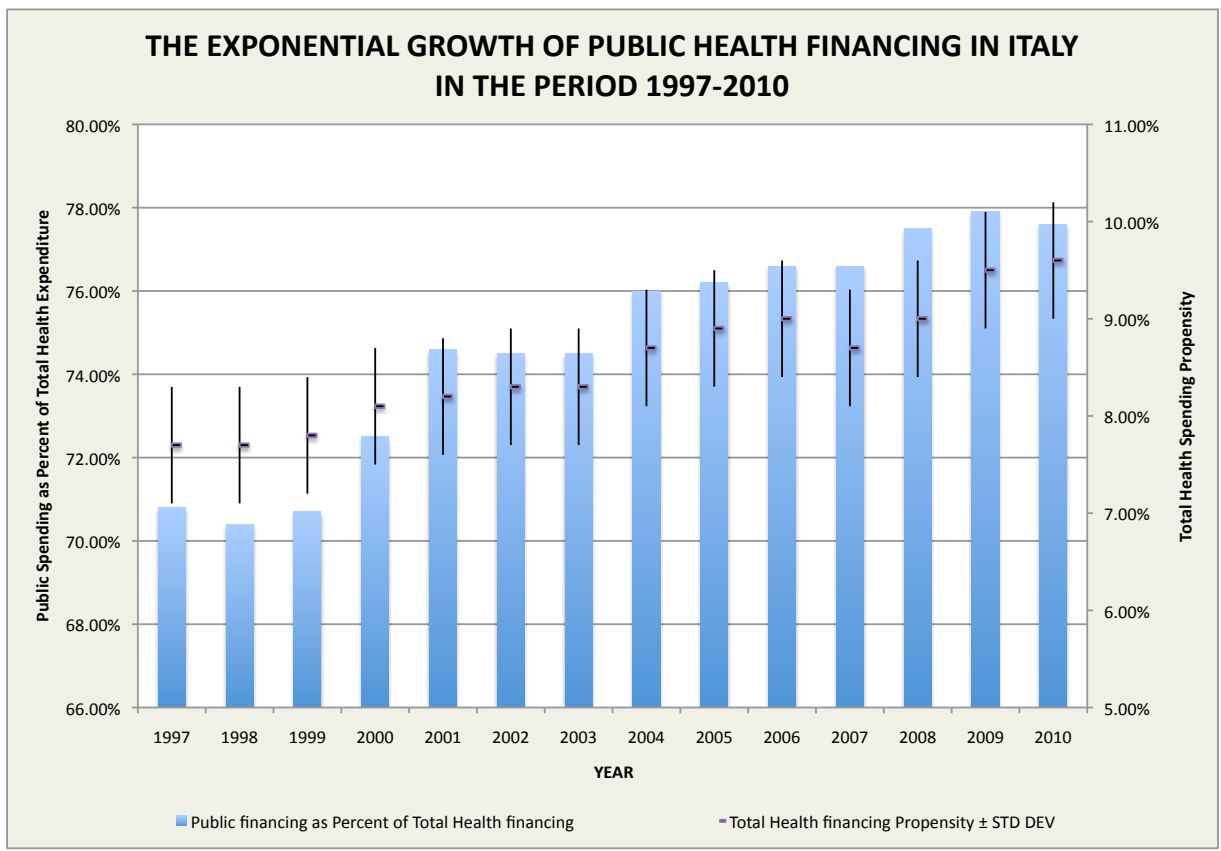

Source: Authors' elaborations on data from OECD HEALTH DATA 201

1.5. *Cæteris paribus* projections[8,9] (Exhibit 2) to the year 2025 and 2050 of the present statistically significant trends highlight an increase of the public health burden to circa 11% and 20% of the GDP respectively[10], a clearly unsustainable trend, even more so in the light of the current depressive Italian GDP long-term growth expectations (Ministero dell'Economia e delle Finanze 2012), the Eurozone financial crisis (OECD 2012) and the prospective growth of the interest service of Italian public debt and of the sovereign debt and deficit containment measures required in all advanced G-20 economies (OECD 2012). In addition, the percent of public health financing on total health financing would reach 85% by the year 2025 and exceed 92% by the year 2050. It therefore quite clear that, regardless of any contingency plans, a structural financial turnaround of the exponential growth rate of public health financing is necessary.

---

[8] For an example of the methodology of Financing growth projection see: Chernew. M e al, *Increased Financing on Health Care: Long-Term Implications for the Nation*, Vol. 28 (Health Affairs, 2009).

[9] For a more specific analysis of the specific drivers of health Financing see: OECD, *Projecting OECD health and long-term care Financings: what are the main drivers?*, Vol. 477 (Paris: Economic Department Papers, 2006)..

[10] As a purely indicative term of comparison, total health Financing in percent of GDP is, in the year 2009, 11.6% in Germany and 17.4% in the USA. The percent of public Financing is 76.9% and 47.7% respectively, making the burden of public health Financing 8.9% in Germany and 8.3% in the USA. http://stats.oecd.org/Index.aspx?DataSetCode=SHA (accessed 8/18/09)

1.6. We will propose in the next chapters a simple analytical model and a synthetic index for the assessment of the *a priori* sustainability of a budgetary intervention on the growth rate of public health financing. In particular we will analyze the Italian *Documento di Economia e Finanza 2011* (Ministero dell'Economia e delle Finanze 2011)[11] for the period 2011-2014.

## 2. The Rationale Behind the Sustainability of a Public Health Intervention

2.1. The sheer amount of public health spending and of public health spending growth considered quantitatively adequate and sustainable for an economy is not an absolute, but a *relative* concept which needs to be analyzed within a *dynamic framework*, where the contingent tradeoffs among the *endogenous* and *exogenous* financial, economic, demographic, epidemiological and social variables determining *sustainability* must be assessed (Chunling Lu 2010). In this paper we will focus on the financial sustainability of a health system guaranteeing universal coverage heading into the Eurozone financial crisis.

2.2. According to the World Health Organization (Thomson, Foubister and Mossialos 2009), the Sustainability[12] of a health system is determined by its Economic and Fiscal Sustainability.

2.3. Economic Sustainability specifically refers to growth in public health financing as a proportion of gross domestic product (GDP). Financing on health is sustainable up to the point at which the social cost of health financing exceeds the value produced by that financing. If health financing sufficiently threatens other valued areas of economic activity, health financing may come to be seen as economically unsustainable. In order to exemplify, every Euro spent on health care represents one fewer Euro spent on education, national defense, housing, subsidies. The more we spend on health care, the less we are able to spend elsewhere (Thomson, Foubister and Mossialos 2009).

---

[11] www.mef.gov.it, www.dt.tesoro.it ,www.rgs.tesoro.it (accessed 8/18/09)
[12] The term actually used by S Thomson, T Foubister and E Mossialos, *Financing Health Care in the European Union*, ed. European Observatory on Health Systems and Policies Studies Series N. 17 (World Health Organization, 2009) is Economic Sustainability, but in the present model we have split the original meaning into Financial and Economic Sustainability, where, for the purpose of the present paper, the former conforms the best to the original definition.

2.4. Fiscal Sustainability[13] of a health system relates specifically to public financing on health care. A health care system may be economically sustainable and yet fiscally unsustainable if internal public revenue is not sufficient to meet public financing (Thomson, Foubister and Mossialos 2009).

2.5. However, we consider this approach to sustainability not fully satisfactory when the dynamic uncertainty of an economic downturn and an aging population (Nanako Tamiya 2011), such as is experiencing Italy in 2011 and 2012, in a wider global scenario, could both suffer and produce wide fluctuations, if not a decline, in the GDP and, consequently, a risk of reduction in public health financing. In Italy, in particular (Exhibit 1), such reduction in public financing could have the effect of increasing the out-of-pocket health financing which, in times of crisis, could hamper universal coverage with adverse epidemiological[14] consequences on the old (Kenneally M 2012) and less well-to-do households (Kenji Shibuya 2011). In addition, when a moderate albeit insufficient growth of GDP is present, non-health financing, such as the growing interest service on sovereign debt, deficit and debt reduction and growing unemployment subsidies, could require health applications to decline in favour of growing non-health applications. Therefore if, on one hand the health regulator is obliged to increase its public health financing propensity when the GDP is decreasing in order to maintain public health expenditure unvaried in order to prevent social and epidemiological instability deriving from inequality in the constitutional right of universal health coverage, on the other it is obliged again to *decrease* its public health care financing propensity when the GDP is growing in favour of an allocation of funds to deficit and sovereign debt reduction interventions (OECD 2011).

---

[13] In this paper we will analyze only the Economic Sustainability of the Ministero dell'Economia e delle Finanze, *Documento di Economia e Finanza 2011* (Roma: Ministero dell'Economia e delle Finanze, 2011).

[14] The epidemiological effects of the financing model of a health system is the object of several researches on the relationship between life expectancy and the cost of health care in the United States. In particular, the Organization for Economic Cooperation and Development (OECD) issued in December 2008 an Assessment and Recommendations Policy Brief stating that Health Care Reform is needed in the US OECD, *Economic Survey of the United States 2008* (Paris: OECD, 2008). The assessment states that despite health spending being much higher in the US (about 15% of GDP) than in any other OECD country and the use of cutting edge technology, the US population's health status does not compare favorably on key indicators. The US ranks poorly in terms of life expectancy at birth, infant mortality and amenable mortality (i.e. mortality that can be averted by good health care).

2.6. In the wider rationale of the model presented in this paper the Sustainability of a Health System is determined by its Social, Epidemiological, Financial, Economic and Fiscal Sustainability.

2.7. The Social Sustainability of a health system pertains the fulfilment of the shared values that the nation holds. In Italy, the Health Reform of 1992-1999 placed the State as guarantor of the equality, solidarity and universal coverage in the provision of health services (Cesana 2005). There follows that any public policy intervention undermining the constitutional principles of equality, solidarity and universal coverage through the outright reduction of public health services is not considered socially sustainable in the present paper .

2.8. The Epidemiological Sustainability of a health system refers to its capacity to meet the global health demands of the population in terms of increasing life expectancy at birth (Kaplan 2009).

2.9. Economic Sustainability is defined in the present paper in strict connection with Social Sustainability, i.e. we regard any public policy intervention which aims at reducing public health resources in absolute terms socially *and* economically sustainable *if* the resources reduced are those resources that are actually *wasted*, i.e. financed and actually expended but *not* utilized efficiently and/or effectively (Pagano and Vittadini 2004) for a valuable (Porter 2009) delivery of care. Therefore Economic Sustainability acts like a bridge between Financial and Social and Epidemiological Sustainability: it makes savings in financial resources socially sustainable through the improvement in the value of the health services provided.

2.10. Following the preceding definitions, we will consider the health system Financially Sustainable if it has the capacity, through increasing efficiency, effectiveness (Pagano and Vittadini 2004) and value per euro spent (Porter 2009), to *decrease* its public health financing propensity when the GDP is growing, i.e. to *freeze* its public health financing growth *and* at the same time meet the social and epidemiological needs of quality and equality in the provision of health care.

2.11. In the model we will focus on the effects of *exogenous* variation on the sustainability of public health universal coverage - in particular on the effects of growing interest rates on sovereign debt - and not in the effects of *endogenous* variation, such as a reduction in military spending. The reason is that the latter, among other expenditures, pertains to the constitutional values a nation holds – arms, pensions or health?[15] The only *endogenous* variable which is addressed in this paper is *waste*, as shall be explained further on.

2.12. Quite simply, in formal terms, if: *P* is the percent of total health expenditure publicly financed; *Pϕ* is the propensity (as percent of the GDP) to spend economic resources in publicly financed health care; *Ω* is (as percent of the GDP) the percent of health resources actually financed but economically *wasted*, i.e. not utilized efficiently and/or effectively for the delivery of health care; $\hat{\Gamma}$ is the *endogenously* determined volume of financial resources utilized for non-health publicly financed applications; *S* the *exogenously* determined interests paid for the service of sovereign debt (as percent of the GDP)[16]; and *T* the total expendable public revenue (as percent of the GDP)[17], then

**Equation 1**

$$TY - (P\varphi - \Omega)Y - SY - \hat{\Gamma} = 0$$

it is clear that, in the present model, public health expenditure (*Pϕ-Ω* )*Y* can increase only if:

a. *Y* increases (as has been the case in Italy in the period 1997-2010 (Exhibit 2));

b. *P* increases (as has been the case in Italy in the period 1997-2010 (Exhibit 1));

c. *Pϕ* increases, and consequently (*T-Pϕ*) decreases (as has been the case in Italy in the period 1997-2010 (Exhibit 2));

d. T increases[18];

e. *SY* and Γ decrease;

d. Ω (*waste*) decreases, i.e. the value (Porter 2009), the efficiency and/or the effectiveness (Pagano and Vittadini 2004) of the health services increases[19].

2.13. For the purpose of this paper we shall evaluate the financial sustainability of a public health intervention not in absolute but in prospective relative terms, i.e. considering the last actual health financing as *de facto* financially, economically, fiscally, epidemiologically and socially sustainable, and analyzing the dynamics of future budgetary interventions. Therefore we will focus the evaluation of the financial sustainability of a budgetary intervention on the variation in the public health financing propensity ($\phi=HY^{-1}$) [Appendix B] with respect both to variation of the GDP (*Y*) *and* to variation of the percent of health spending publicly financed (*P*). Variation in Economic Sustainability is not explicitly analyzed and considered *implicit* in the variation of the public health financing propensity ($P\phi$). In fact, from a strictly *financial* point of view, a reduction in the *waste* ($\Omega$) of resources increases financial sustainability *only* when *both* public health financing and public health financing propensity are stable or reduced, or, in other words, exponential financial growth does not absorb the resources generated by growing effectiveness and/or efficiency. In the present model, health care delivery can be efficient and effective, but nonetheless financially unsustainable.

---

[15] Of course, in the case of an invasion, growing arms expenditure ought to be considered an *exogenous* variation affecting public health spending sustainability

[16] We will focus on interest rates, among the many non-health expenditures, since at the time this paper is being submitted (July-August 2012), Italy and Spain are under sever pressure on growing sovereign debt interest rates.

[17] Current deficit and sovereign debt increase have not been analyzed in the present paper.

[18] Fiscal revenue has not been analyzed in the present paper.

[19] The parameter Ω has not been analyzed in the present paper.

2.14. In partial differential terms:

**Equation 2**

$$\text{Financial Sustainability } (\sigma) = \frac{\partial^2 (PHY^{-1})}{\partial P \partial Y} \frac{PY}{PHY^{-1}} = \frac{\partial}{\partial P}\left(-PHY^{-2}\right)\frac{Y}{HY^{-1}} = -1$$

and in absolute differential terms:

**Equation 3**

$$\text{Financial Sustainability } (\sigma) = \frac{\Delta P \cdot HY^{-1} + \Delta(HY^{-1}) \cdot P + \Delta P \Delta(HY^{-1})}{\Delta P \cdot Y + \Delta Y \cdot P + \Delta P \Delta Y} \frac{Y}{HY^{-1}}$$

where [Appendix B]:

**Definition 1**

$$\text{Financial Sustainability } (\sigma) = \begin{cases} \ll -1; & \text{reduction of public health financing - social and epidemiological} \\ & \text{sustainability at risk if economic sustainability does not increase;} \\ \sigma = -1; & \text{stabilization of public health spending;} \\ -1 < \sigma < 0; & \text{range of intervention financially sustainable;} \\ = 0; & \text{stabilization of public health spending propensity growth;} \\ > 0; & \text{lower limit of financial unsustainability;} \\ \gg 0; & \text{financial and fiscal sustainability at risk.} \end{cases}$$

2.15. In particular, we will analyze the *Documento di Economia e Finanza 2011 – Programma Nazionale di Riforma* where, as we shall see in detail in the future chapter, public health financing propensity growth is stabilized, which in analytic terms, means that public health financing will continue to grow only if and at the same rate of the growth of the GDP (Ministero dell'Economia e delle Finanze 2011), i.e. *both* the variation in the percent of public health financing on the total health care financing *and* the variation in the public health care financing propensity are null.

# 3. An Analysis of the Italian Stability and National Reform Program of the Documento di Economia e Finanza 2011 on Public Health Financing

3.1. The *Documento di Economia e Finanza 2011* (DEF 2011)[20] contains several containment measures of the growth of public health financing for the period 2011-2014. Such measures are outlined in *§ III.3 SANITA'* of *Sezione II: Analisi e Tendenze della Finanza Pubblica* and *§ IV.3 LA SPESA SANITARIA* of *Sezione II: Nota Metodologica (Allegato)*.

3.2. Data relevant to the present analysis are summarized in Exhibit 4[21,22]

---

[20] www.mef.gov.it, www.dt.tesoro.it ,www.rgs.tesoro.it

[21] The model highlights a discrepancy in the actual 2010 data between OECD HEALTH DATA 2011 and DEF 2011at . However such discrepancy is in the order of 5 base points in the exponential growth rates, and therefore not significant.

[22] Exhibit 4 has been deliberately left in Italian for an easier cross-reference with the original public accounting document which can be found at [22] www.mef.gov.it, www.dt.tesoro.it ,www.rgs.tesoro.it
The variables of interest have been highlighted.

**Exhibit 4**

| DOCUMENTO DI ECONOMIA E FINANZA 2011<br>ANALISI E TENDENZE DELLA FINANZA PUBBLICA<br>CONTO ECONOMICO DELLE AMMINISTRAZIONI PUBBLICHE | | | | | | | 2010-2014 | |
|---|---|---|---|---|---|---|---|---|
| TABELLA II.2-1 | | 2010 | 2011 | 2012 | 2013 | 2014 | CORRELATION COEFFICIENT | LOG REGRESSION |
| **SPESE** | | | | | | | | |
| Redditi da lavoro dipendente | | 171,905 | 171,090 | 170,693 | 170,840 | 172,191 | | |
| Consumi intermedi | | 137,009 | 137,425 | 138,857 | 142,366 | 147,081 | | |
| Prestazioni sociali | | 298,199 | 306,200 | 313,630 | 324,940 | 336,540 | | |
| Pensioni | | 236,931 | 244,630 | 252,100 | 260,790 | 270,740 | | |
| Altre prestazioni sociali | | 61,268 | 61,570 | 61,530 | 64,150 | 65,800 | | |
| Altre uscite correnti | | 62,349 | 62,392 | 60,622 | 61,029 | 61,416 | | |
| **Totale spese correnti netto interessi** | | **669,462** | **677,107** | **683,802** | **699,175** | **717,228** | | |
| Interessi passivi | SY | 70,152 | 76,087 | 84,023 | 91,313 | 97,605 | 1.00 | 8.79% |
| **Totale spese correnti** | | **739,614** | **753,194** | **767,825** | **790,488** | **814,833** | | |
| *di cui: Spesa sanitaria* | P$\varphi$Y | 113,457 | 114,836 | 117,391 | 122,102 | 126,512 | 0.98 | 2.83% |
| **Totale spese in conto capitale** | | **53,899** | **48,691** | **45,217** | **46,037** | **45,956** | | |
| Investimenti fissi lordi | | 31,879 | 31,230 | 27,014 | 27,816 | 28,192 | | |
| Contributi in c/capitale | | 20,442 | 17,826 | 16,058 | 16,109 | 16,104 | | |
| Altri trasferimenti | | 1,578 | -365 | 2,145 | 2,112 | 1,660 | | |
| **Totale spese netto interessi** | | **723,361** | **725,798** | **729,019** | **745,212** | **763,184** | | |
| **Totale spese finali** | | **793,513** | **801,885** | **813,042** | **836,525** | **860,789** | | |
| | | | | | | | | |
| **ENTRATE** | | | | | | | | |
| Tributarie | | 445,416 | 457,066 | 476,544 | 492,008 | 507,935 | | |
| Imposte dirette | | 225,494 | 230,221 | 242,320 | 250,379 | 257,940 | | |
| Imposte indirette | | 216,530 | 226,272 | 233,645 | 241,043 | 249,401 | | |
| Imposte in c/capitale | | 3,392 | 573 | 579 | 586 | 594 | | |
| Contributi sociali | | 214,508 | 219,820 | 225,447 | 230,813 | 237,360 | | |
| Contributi sociali effettivi | | 210,460 | 215,701 | 221,267 | 226,574 | 233,060 | | |
| Contributi sociali figurativi | | 4,048 | 4,119 | 4,180 | 4,239 | 4,300 | | |
| Altre entrate correnti | | 58,583 | 58,472 | 60,513 | 61,948 | 63,536 | | |
| **Totale entrate correnti** | | **715,115** | **734,785** | **761,925** | **784,183** | **808,237** | | |
| Entrate in conto capitale non tributarie | | 3,795 | 4,608 | 5,678 | 5,998 | 6,069 | | |
| **Totale entrate finali** | TY | **722,302** | **739,966** | **768,182** | **790,767** | **814,900** | 1.00 | 3.12% |
| *Pressione fiscale* | | 42.6% | 42.5% | 42.7% | 42.6% | 42.5% | | |
| | | | | | | | | |
| **Saldo primario** | | **-1,059** | **14,168** | **39,163** | **45,555** | **51,716** | | |
| **Saldo di parte corrente** | | **-24,499** | **-18,409** | **-5,900** | **-6,305** | **-6,596** | | |
| **Indebitamento netto** | | **-71,211** | **-61,919** | **-44,860** | **-45,758** | **-45,889** | | |
| | | | | | | | | |
| **PIL nominale** | Y | **1,548,816** | **1,593,314** | **1,642,432** | **1,696,995** | **1,755,013** | **1.00** | **3.18%** |

*SOURCE: MINISTERO DELL'ECONOMIE E DELLE FINANZE: DEF 2011, TABELLA II.2-1*

It is quite clear that the economic and financial stability plan is based on the assumption that the GDP will grow at a rate of 3.18% (Correlation Coefficient 1.00) in the period 2010-2014[23].

3.3. In more detail, for the period 2010 (actual) – 2014 (plan) (Ministero dell'Economia e delle Finanze 2011), the coefficients (See Exhibit 4 and Appendix A) of Equation 1 are summarized in Exhibit 5:

---

[23] As of September 1, 2011, see the analyses of a possible stagnation in www.istat.it, the €-coin index in www.bancaditalia.it and OECD, *RAPPORTO ITALIA 2011* (OECD - Ministero dell'Economia (Traduzione a cura di), 2011).

**Exhibit 5**

| DOCUMENTO DI ECONOMIA E FINANZA 2011<br>ANALISI E TENDENZE DELLA FINANZA PUBBLICA<br>TABLE OF THE COEFFICIENTS | | | | | |
|---|---|---|---|---|---|
| | 2010 | 2011 | 2012 | 2013 | 2014 |
| T | 46.64% | 46.44% | 46.77% | 46.60% | 46.43% |
| P | 77.60% | 77.60% | 77.60% | 77.60% | 77.60% |
| φ | 9.44% | 9.29% | 9.21% | 9.27% | 9.29% |
| Pφ | 7.33% | 7.21% | 7.15% | 7.20% | 7.21% |
| S | 4.53% | 4.78% | 5.12% | 5.38% | 5.56% |
| (T−S)−Pφ | 34.78% | 34.46% | 34.51% | 34.02% | 33.66% |

*SOURCE: MINISTERO DELL'ECONOMIE E DELLE FINANZE: DEF 2011, TABELLA II.2-1*

3.4. Without unnecessary statistical sophistication, Exhibits 4 & 5 highlight that, in the period 2010-2014, the Italian government plans to *freeze* : a) the growth of public health financing propensity $P\phi$ but *not* the growth of public health financing $P\phi Y$ (exp=2.83%, Correlation Coefficient 0.98); b) the growth of public revenue T; and c) to reduce non-health applications ($T$-$P\phi$) in favour of the growing burden of the service of sovereign debt ($S$).

3.5. In the wording of the rationale of the present model, it is also quite clear that, if the GDP $Y$ growth does not meet the expectations (increased Financial Sustainability) and/or interventions towards value, efficiency and/or effectiveness do not reduce the waste of resources $\Omega$ (increased Economic Sustainability), either fiscal pressure $T$ must grow (decreasing Fiscal Sustainability) or health applications $P\phi$ decrease (decreasing Social and Epidemiological Sustainability).

3.6. In absolute terms, in the DEF 2011 (Ministero dell'Economia e delle Finanze 2011), the Average Growth [Appendix A] of public health financing in the period 2011-2014 amounts to circa 4,000 million euro per year (Exhibit 6). In the same period, average growth will exceed Cumulated Marginal and Average Growth [Appendix A], which, in micro-analytic terms, means that the Italian DEF 2011 aims to reduce the growth rate of public health financing (Exhibit 7).

**Exhibit 6**

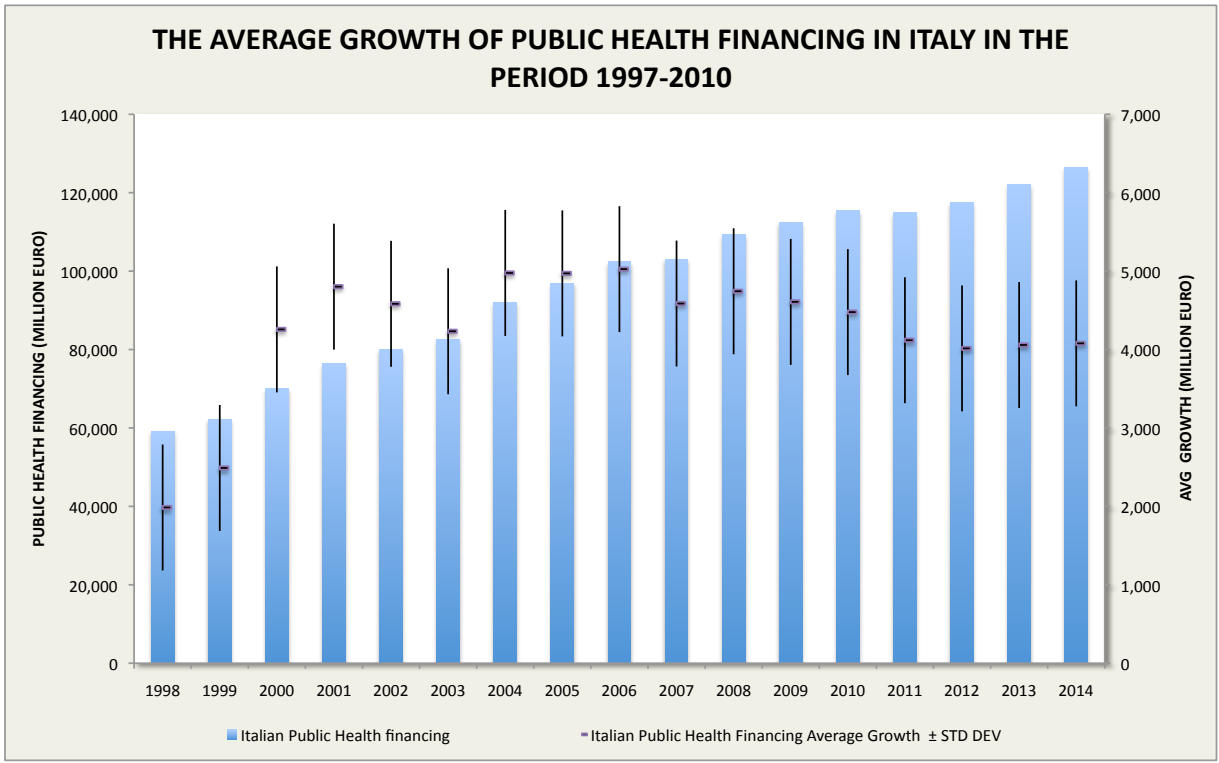

Source: Authors' elaborations on data from OECD HEALTH DATA 2011 and MINEF: Documento di Economia e Finanza 2011

**Exhibit 7**

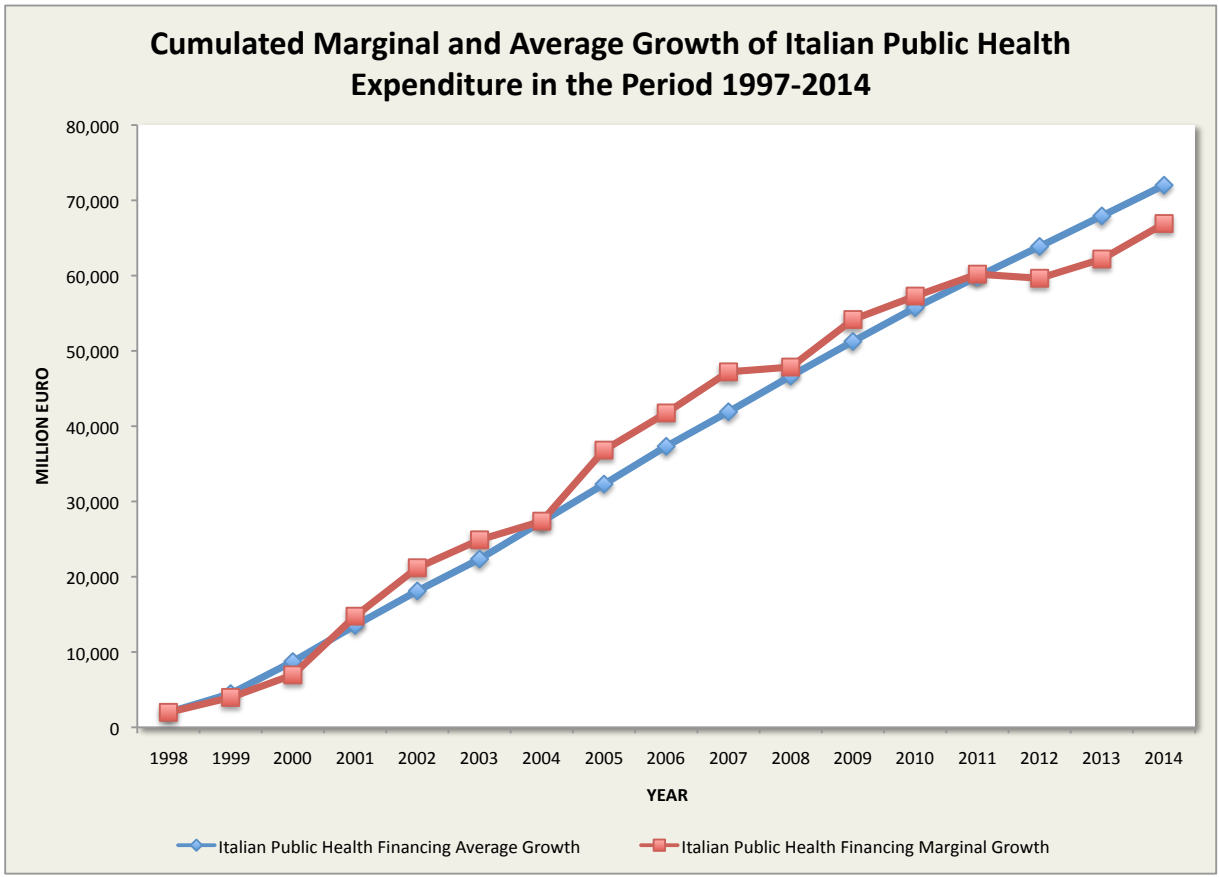

Source: Authors' elaborations on data from OECD HEALTH DATA 2011 and MINEF: Documento di Economia e Finanza 2011

3.7. However, absolute marginal and average growth representation (Blume 2007) does not analyze the dynamic relationships between the variables which determine the financial sustainability of the public intervention plan. In Appendices A and B we have developed a synthetic analytical model for the assessment of the effectiveness of such intervention, whose results are shown in Exhibit 7. In synthesis, the DEF 2011 is based on the assumptions that, in the period 2011-2014:

3.7.1. the GDP ($Y$) will grow at a rate of 3.18% (Correlation Coefficient 1.0), i.e. at almost the same rate of 3.23% (Correlation Coefficient 0.98) of the period 1997-2010;

3.7.2. the growth of the public percent of the total health financing (P) is null, vs. the growth of 0.82% (Correlation Coefficient 0.96) in the period 1997-2010; On the basis of these assumptions, the working hypothesis is that:

3.7.3. the growth of the health financing propensity ($P\phi$) will be stabilized (Exponential Growth Rate -0.67%, Correlation Coefficient -0.66);

3.7.4. public health expenditure will continue to grow at a rate of 2.83% (Correlation Coefficient 0.98), vs. the rate of 5.74% (Correlation Coefficient 0.99) in the period 1997-2010.

**Exhibit 8**

| THE FINANCIAL SUSTAINABILITY OF THE ITALIAN DEF 2011 | | OECD ACTUAL | | 1997-2010 | | DFP 2011 | | | 2010-2014 | | |
|---|---|---|---|---|---|---|---|---|---|---|---|
| | YEAR | 1997 | 2010 | GROWTH AS A FUNCTION OF TIME | | 2011 | 2012 | 2013 | 2014 | GROWTH AS A FUNCTION OF TIME | |
| | PERIODS | | 0 | CORRELATION COEFFICIENT | EXPONENTIAL GROWTH RATE - LOG REGRESSION | 1 | 2 | 3 | 4 | CORRELATION COEFFICIENT | EXPONENTIAL GROWTH RATE - LOG REGRESSION |
| Public financing as Percent of Total Health financing | P | 70.80% | 77.60% | 0.96 | 0.82% | 77.60% | 77.60% | 77.60% | 77.60% | | 0.00% |
| Italian Public Health financing | PH | 57,175 | 115,381 | 0.99 | 5.74% | 114,836 | 117,391 | 122,102 | 126,512 | 0.94 | 2.46% |
| Italian Public Health financing Propensity | Pφ | 5.45% | 7.45% | 0.98 | 2.51% | 7.21% | 7.15% | 7.20% | 7.21% | -0.66 | -0.67% |
| Nominal Gross Domestic Product | Y | 1,048,766 | 1,548,816 | 0.98 | 3.23% | 1,593,314 | 1,642,432 | 1,696,995 | 1,755,013 | 1.00 | 3.13% |
| *FINANCIAL SUSTAINABILITY* | σ | | 0.59 | | | -1.13 | -0.27 | 0.20 | 0.05 | | |

*SOURCE: Authors' elaborations on OECD HEALTH DATA 2011 and MINEF: DOCUMENTO DI ECONOMIA E FINANZA 2011*

3.8. The result of the budgetary intervention (Exhibits 7 & 8) is that the Sustainability Coefficient ($\sigma$) [Appendix B], whose average value has been 0.59 in the period 1997-2010, and therefore in the financial sustainability range ($0<\sigma<1$), has been reduced to -1.13 in 2011 in the conditional social end epidemiological unsustainability risk range ($\sigma<-1$), only to grow again and fluctuate around the zero value (-0.27 in 2012, 0.20 in 2013 and 0.05 in 2014), thus highlighting the intention of the *Ministero dell'Economia* to stabilize public health financing growth but *not* to *freeze* it in favour of non-health applications. In analytic terms, in the year 2011, with ΔP=0 and Δφ decreasing:

$$\text{Financial Sustainability } (\sigma) = \left| \frac{\Delta(HY^{-1}) \cdot P}{\Delta Y \cdot P} \frac{Y}{HY^{-1}} \right|_{\delta P \to 0} = -1.13$$

In the years 2012, 2013 and 2014, if ΔP=0 and Δφ≈0:

$$\text{Financial Sustainability } (\sigma) = \left| \frac{\succ 0}{\Delta Y \cdot P} \frac{Y}{HY^{-1}} \right|_{\substack{\delta P \to 0 \\ \delta HY^{-1} \to 0}} =$$
$$= -0.27 (2012); \ 0.20 \ (2013); \ 0.05 (2014)$$

3.9. Remarking that it is beyond the scope of this paper: a) to predict the growth rate of the GDP in the period 2011-2014; b) to assess the probability that the exponential growth of the public percent of the total health bill will be null, even in the face of the diffuse social effects of the economic downturn, expecially in the south of Italy; and c) to assess the international demands towards financial non-health applications such as sovereign debt and deficit reduction - we observe here that, in the case GDP growth is very small in the period 2011-2014 (OECD 2011), Equation 4 would converge towards σ≈1 (financial and fiscal unsustainability). In analytic terms:

$$\text{Financial Sustainability } (\sigma) = \left| \frac{\Delta P \cdot HY^{-1}}{\Delta P \cdot Y} \frac{Y}{HY^{-1}} \right|_{\substack{\delta HY^{-1} \to 0 \\ \delta Y \to 0}} \simeq 1$$

3.10. In intuitive graphic form, Exhibit 8 shows quite clearly that, in the period 2011-2014, the Economic Sustainability Coefficient oscillates between -1 and 0, thus highlighting the effort of the italian government to stabilize at first public health financing propensity (σ<-1 in 2011 ), at the risk of social unsustainability if efficiency and effectiveness in the delivery of care do not increase proportionally, and then conditioning the growth of public health financing (σ≈0 in 2012, 2013 and 2014) to that of the GDP.

**Exhibit 9**

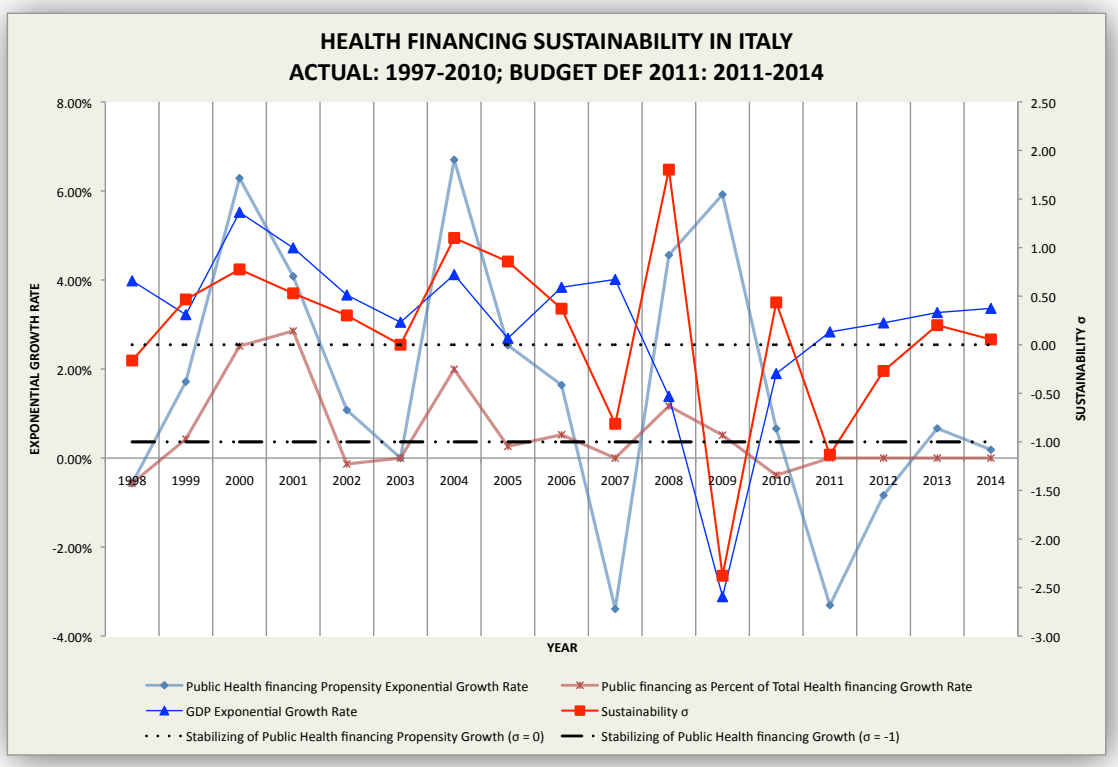

Source: Authors' elaborations on data from OECD HEALTH DATA 2011 and MINEF: Documento di Economia e Finanza 2011

3.11. In Exhibit 8 we remark again that it is also quite evident that most of the working hypothesis rests on the assumption that the GDP will grow again at an exponential rate beginning in the year 2011, and that non-health applications will be stable in the period 2011-2014. Such working hypotheses ought to be reviewed in the light of recent forecasts on italian and global economic growth[24] .

# 4. Conclusions

The rationale behind the analytic model presented in this paper highlights the potential trade-offs in the pursuit of social, epidemiological, financial, economic and fiscal sustainability when health financing and health financing propensity stabilization and/or reduction are not accompanied by non-contingent structural reforms towards higher efficiency, effectiveness and value-added in the provision of health services.

---

[24] As of September 1, 2011, see the analyses of a possible stagnation in www.istat.it, the €-coin index in www.bancaditalia.it and OECD, *RAPPORTO ITALIA 2011* (OECD - Ministero dell'Economia (Traduzione a cura di), 2011).

In particular, the Italian *Documento di Economia e Finanza 2011* of the *Ministero dell'Economia e delle Finanze* focuses its intervention, as far as public health care financing is concerned, on the *freezing* of the exponential growth of public health financing propensity (circa 7% of the GDP) while, at the same time, parameterizing the growth rate of public health financing (exp=2.83%, Correlation Coeficient 0.98) to that of the Gross Domestic Product (exp=3.18%, Correlation Coeficient 1.00).

In absolute terms, the average growth of public health financing will be of circa 4,000 billion euro per year in the period 2011-2014, lower than the marginal growth rate in the same period.

The model presented in this paper analyzes the dynamic relationships between the financial variables into a synthetic coefficient.

This intervention of the DEF 2011 is *a priori* Financially Sustainable with a Sustainability Coefficient fluctuating around zero (Upper limit of Financial Sustainability). This working hypothesis rests on the assumptions that the Gross Domestic Product will grow in the period 2011-2014 exponentially at a rate of 3.18%, vs. a growth of 3.29% in the period 1997-2010, that the public component of health care financing will not grow in the period 2011-2014, vs. a growth of 0.82% in the period 1997-2010 and that non-health applications, such as sovereign debt service and deficit reduction, will not be affecting health applications.

In case the growth of the GDP in the period 2011-2014 will be small (OECD 2011), the Financial Sustainability of the intervention would be in jeopardy, and the Sustainability Coefficient would grow above 1 in the Financial and Fiscal Unsustainability range.

Further analysis should asses the effects on the social, epidemiological and economic sustainability of the present intervention, in particular it should assess the hypothesis that the *freezing* of public health financing propensity growth, not accompanied by increasing economic sustainability, does not increase private out-of-pocket and insured health financing propensity, thus partially contradicting the fundamentals of the Italian Health Reform of 1992-1999.

# Appendix A : Health Financing Propensity Average and Marginal Growth

The present Appendix A summarizes some well known standard mathematical expressions of the Exponential Growth Rate of Health Financing Propensity and its Average and Marginal Growth *with the only scope of clarifying the utilization of the variables of the model*. *The reading of this Appendix A should be omitted by researchers already familiar with basic mathematical dynamic variables analysis*.

A.1. For the purpose of the present analysis we shall define:

t = time in terms of number of periods;

Y = the nominal gross domestic product (GDP);

g = the exponential growth rate of Y;

H = the nominal total health care Financing;

h = the exponential growth rate of H;

ϕ = health care Financing propensity;

f = the exponential growth rate of ϕ;

P = the public percent of the total health care Financing H;

p = the exponential growth rate of P;

O = the out-of-pocket percent of the total health care Financing H;

o = the exponential growth rate of O;

I = the privately insured percent of the total health care Financing H;

i = the exponential growth rate of I;

where:

**Equation 4**

$$P_0 e^{pt} + O_0 e^{ot} + I_0 e^{it} = 1$$

A.2. OECD HEALTH DATA 2011 datasets show the incidence of health Financing as percent of the gross domestic product expressed in static form. The data can be described in the form of the coefficient:

**Equation 5**

$$\varphi = HY^{-1}$$

where *H* and *Y* are independent variables.

A.3. In the present paper we will analyze the dynamic evolution of *H* with respect to time, in the form of the relationship:

**Equation 6**

$$\frac{dH}{dt} = hH$$

which is a possible solution of

$$H = H_0 e^{ht}$$

in fact

$$\frac{d}{dt}\left[H_0 e^{ht}\right] = H_0 e^{ht} h = Hh$$

Since we can also express H as a function of (Y, g, φ, f) in the explicit dynamic form:

**Equation 7**

$$H = \varphi_0 e^{ft} Y_0 e^{gt} = \varphi_0 Y_0 e^{ft} e^{gt} = H_0 e^{(f+g)t}$$

where

**Equation 8**

$$\frac{dH}{dt} = \frac{d}{dt}\left(\varphi_0 e^{ft} Y_0 e^{gt}\right) = \varphi_0 Y_0 (e^{ft} f e^{gt} + e^{ft} e^{gt} g) = H_0 e^{(f+g)t}(f+g) = H(f+g)$$

equating [2] and [4]:

$$Hh = H(f+g)$$

and

**Equation 9**

$$h = f + g$$

therefore [1] will be expressed as

$$f = h - g$$

A.4. As far as *only* public health financing will be analyzed, we will express [4] as:

**Equation 10**

$$H = P_0 \varphi_0 e^{(p+f)t} Y_0 e^{gt} = P_0 H_0 e^{[(p+f+g)]t} = P_0 H_0 e^{[(p+h)]t}$$

and [6] as

**Equation 11**

$$h + p = p + f + g$$

A.5. Equation [5] is analogous to OECD HEALTH DATA 2011 Equation [1], but in dynamic form, where $(p+f)$ is the exponential growth rate of the propensity Pφ of the system to spend its resources in public health care. Therefore projections for *t* periods, linear Correlation Coefficients and Logarithmic Regressions (Glantz and Slinker 2001) are derived from datasets in the form:

$$\ln H_t = \ln H_0 + [p + f + g]t$$

and

**Equation 12**

$$f = \frac{1}{t}(\ln H_t - \ln H_0) - g - p$$

A.6. We have already defined the growth of public health financing as a function of time. We will define the average growth of public health financing as (Blume 2007):

**Equation 13**

$$AG(t) = \frac{P(t)H(t)}{t}$$

and marginal growth:

**Equation 14**

$$MG(t) = \frac{d[P(t)H(t)]}{dt}$$

where:

$$\frac{d}{dt}AG(t) = \frac{d}{dt}\left[\frac{P(t)H(t)}{t}\right] = \frac{\frac{d[P(t)H(t)]}{dt}t - P(t)H(t)}{t^2} =$$

$$= \frac{\frac{d[P(t)H(t)]}{dt} - \frac{P(t)H(t)}{t}}{t} = \frac{MG - AG}{t}$$

# Appendix B: Financial Sustainability

B.1. We will postulate that the growth/decline of public health financing is financially sustainable if the relative growth/decline of its public health spending propensity is perfectly elastic relatively to the relative growth/decline of the nominal GDP of the system *and* to the relative growth/decline of the growth rate of the percent of health spending publicly financed.

B.2. In partial differential form:

**Equation 15**

$$\text{Financial Sustainability } (\sigma) = \frac{\partial^2 (PHY^{-1})}{\partial P \partial Y} \frac{PY}{PHY^{-1}} = \frac{\partial}{\partial P}\left(-PHY^{-2}\right)\frac{Y}{HY^{-1}} = -1$$

B.3. In absolute differential form:

**Equation 16**

$$\text{Financial Sustainability } (\sigma) = \frac{\frac{\Delta(PHY^{-1})}{PHY^{-1}}}{\frac{\Delta(PY)}{PY}} = \frac{\Delta(PHY^{-1})}{\Delta(PY)} \frac{PY}{PHY^{-1}} = \frac{\Delta(PHY^{-1})}{\Delta(PY)} \frac{Y}{HY^{-1}} =$$

$$= \frac{\Delta P \cdot HY^{-1} + \Delta(HY^{-1}) \cdot P + \Delta P \Delta(HY^{-1})}{\Delta P \cdot Y + \Delta Y \cdot P + \Delta P \Delta Y} \frac{Y}{HY^{-1}}$$

where

**Definition 1**

$$\text{Financial Sustainability } (\sigma) = \begin{cases} \ll -1; \text{ reduction of public health financing - social and epidemiological} \\ \quad \text{sustainability at risk if economic sustainability does not increase;} \\ \sigma = -1; \text{ stabilization of public health spending;} \\ -1 < \sigma < 0; \text{ range of intervention financially sustainable;} \\ = 0; \text{ stabilization of public health spending propensity growth;} \\ > 0; \text{ lower limit of financial unsustainability;} \\ \gg 0; \text{ financial and fiscal sustainability at risk.} \end{cases}$$

In fact, substituting in Equation 16:

B.3.1 If the intervention will not increase the percent of health financing publicly financed:

**Equation 17**

$$\text{Financial Sustainability } (\sigma) = \left| \frac{\Delta(HY^{-1}) \cdot P}{\Delta Y \cdot P} \frac{Y}{HY^{-1}} \right|_{\delta P \to 0} \simeq -1$$

which is the base model utilized for the analysis of the Italian *Documento di Economia e Finanza 2011,* since data about the variation in the percent of public financing are not supplied, and, if they were, it would still be contradictory that a public financial containment budgetary intervention plans to increase the percent of expenditure publicly financed, even if in actual terms this is exactly what happened in the period 1997-2010. In this case public health financing propensity is reduced/ (increased) elastically as the GDP increases/(decreases), with the percent of public health financing unchanged. This model guarantees financial *and* social sustainability provided that it has the capacity, through increasing efficiency, effectiveness (Pagano and Vittadini 2004) and value per euro spent (Porter 2009), to *decrease* its public health financing propensity when the GDP is growing, i.e. to *freeze* its public health financing *and* meet the needs of health care and equality.

B.3.2. In the *Documento di Economia e Finanza 2011*, after the year 2011 the regulator specifically aims in the years 2012, 2013 and 2014 at stabilizing the public health financing propensity growth, therefore substituting ΔP≈0 *and* Δϕ≈0 in Equation 15:

**Equation 18**

$$\text{Financial Sustainability } (\sigma) = \left| \frac{\succ 0}{\Delta Y \cdot P} \frac{Y}{HY^{-1}} \right|_{\substack{\delta P \to 0 \\ \delta HY^{-1} \to 0}} \approx 0$$

which is the specific applied outcome of the Italian budgetary intervention in the *Documento di Economia e Finanza 2011* (Exhibits 7,8). In other words, the datasets highlight a compromise solution in which public health financing growth is stabilized in a predetermined public health financing propensity.

B.3.3. In addition, if the future expected growth/decline of the GDP is very small (δY≈0) (OECD 2011):

**Equation 19**

$$\text{Financial Sustainability } (\sigma) = \left| \frac{\Delta P \cdot HY^{-1}}{\Delta P \cdot Y} \frac{Y}{HY^{-1}} \right|_{\substack{\delta HY^{-1} \to 0 \\ \delta Y \to 0}} \simeq 1$$

In this case the exponential growth of public health spending would not be financially sustainable, even if the growth of the public health spending propensity is *frozen*.